\documentclass{aa}\usepackage{graphics,epsf}
\newcommand{\lta}{\;
  \raise0.3ex\hbox{$<$\kern-0.75em\raise-1.1ex\hbox{$\sim$
  }}\;\hskip-2pt }
\newcommand{\gta}{\;
  \raise0.3ex\hbox{$>$\kern-0.75em\raise-1.1ex\hbox{$\sim$
  }}\;\hskip-2pt }
\begin{document}
\thesaurus{06(06.13.1; 02.03.1; 02.13.1)}
\title{Spatiotemporal fragmentation as a mechanism 
for different dynamical modes of behaviour 
in the solar convection zone}
\author{Eurico Covas\thanks{e-mail: eoc@maths.qmw.ac.uk}\inst{1}
\and Reza Tavakol\thanks{e-mail: reza@maths.qmw.ac.uk}\inst{1}
\and David Moss\thanks{e-mail: moss@ma.man.ac.uk}\inst{2}
}
\institute{Astronomy Unit, School of Mathematical Sciences,
Queen Mary and Westfield College, Mile End Road, London E1 4NS, UK
\and Department of Mathematics, The University, Manchester M13 9PL, UK
}
\date{Received ~~ ; accepted ~~ }
\offprints{\em E.\ Covas}
\titlerunning{Covas et al.: Spatiotemporal fragmentation and solar torsional oscillations}
\maketitle
\markboth{Covas et al.: Spatiotemporal fragmentation and solar torsional oscillations}
{Covas et al.: Spatiotemporal fragmentation and solar torsional oscillations}
\begin{abstract}
Recent analyses of the helioseismic
observations
indicate that the previously observed
surface torsional oscillations with periods
of about 11 years
extend significantly  downwards
into the solar convective zone.
Furthermore, there are
indications that the dynamical regimes at the base
of the convection zone are different from those
observed at the top, having either significantly shorter
periods or non--periodic behaviour.

We propose that this behaviour can be explained by the occurrence of
{\it spatiotemporal fragmentation}, a crucial
feature of which is that
such behaviour can be explained solely through nonlinear
spatiotemporal dynamics, without requiring
separate mechanisms with different time scales at
different depths.

We find evidence for this mechanism in the context of 
a two dimensional axisymmetric mean field
dynamo model operating in a spherical shell, with 
a semi--open outer boundary condition,
in which the only nonlinearity  is the action
of the azimuthal component of the Lorentz force of the
dynamo generated magnetic field on the solar
angular velocity.
\end{abstract}


\section{Introduction}
Recent analyses of the helioseismic data, from both the 
Michelson
Doppler Imager (MDI) instrument on board the SOHO
spacecraft (Howe et al.\ 2000a) and the Global Oscillation Network Group (GONG)
project (Antia \& Basu 2000) 
have provided strong evidence which indicates that the earlier observed 
time variations of the differential rotation on the solar
surface, the so called `torsional oscillations' with
periods
of about 11 years (e.g.\ Howard \& LaBonte 1980; Snodgrass, Howard \& Webster 1985;
Kosovichev \& Schou 1997; Schou et al.\ 1998), penetrate into the 
convection zone,
to depths of at least 8 percent in radius.

Furthermore, these data have provided some evidence to
suggest that variations in the differential rotation
are also present around the tachocline at the bottom of the convection
zone (Howe et al.\ 2000b).
An important feature that distinguishes these variations from those 
observed at the upper parts of the convection zone 
is that they possess markedly different modes of behaviour:
either possessing distinctly lower periods 
(of $\sim$$1.3$ years), or being non--periodic (Antia \& Basu 2000).
Clearly, to firmly establish the precise nature of 
these variations, future observations are
required. Whatever the outcome of such observations,
however, both these sets of results point to the 
very interesting possibility that
the variations in the differential rotation
can have different periodicities/behaviours
at different depths in the solar
convection zone.
It is therefore important to ask
whether such different variations 
can in principle occur at different parts of the convection zone and,
if so, what could be the possible mechanism(s) for their production.

The aim of this letter is to suggest that a natural mechanism for the
production of such different dynamical modes of behaviour in the convection
zone is through what we call {\it spatiotemporal fragmentation}, i.e.\
the occurrence of dynamical regimes at (given) values of the control
parameters of the system, which possess different temporal behaviours at
different spatial locations.  This is to be contrasted with the usual
temporal bifurcations, with identical temporal behaviour at each spatial
point, which require changes in parameters to occur.

We find evidence for this mechanism in the context of
a two dimensional axisymmetric mean field
dynamo model in a spherical shell,
with a semi--open outer boundary condition,
in which the only nonlinearity  is the action
of the azimuthal component of the Lorentz force of the
dynamo generated magnetic field on the solar
angular velocity. 
The underlying angular velocity is
chosen to be consistent with the most recent helioseismic data.

In addition to producing 
different dynamical variations in the differential rotation,
including different periods, at the top and 
the bottom of the convection zone, 
this model is also capable of
producing butterfly diagrams which are in qualitative agreement
with the observations as well as displaying
torsional oscillations that penetrate
into the convection zone, as recently  observed by
Howe et al.\ 2000a and Antia \& Basu 2000
and studied by Covas et al.\ (2000).

\section{The model}
We shall assume that the gross features of the 
large scale solar magnetic field 
can be described by a mean field dynamo
model, with the standard equation 
\begin{equation}
\frac{\partial{\bf B}}{\partial t}=\nabla\times({\bf u}\times {\bf B}+\alpha{\bf
B}-\eta\nabla\times{\bf B}).
\label{mfe}
\end{equation}
Here ${\bf u}=v\mathbf{\hat\phi}-\frac{1}{2}\nabla\eta$, 
the term proportional to
$\nabla\eta$ represents the effects of turbulent diamagnetism,
and the velocity field is taken to be of the form $
v=v_0+v'$,
where $v_0=\Omega_0 r \sin\theta$, $\Omega_0$ is a prescribed
underlying rotation law and the component $v'$ satisfies
\begin{equation}
\frac{\partial v'}{\partial t}=\frac{(\nabla\times{\bf B})\times{\bf B}}{\mu_0\rho
r \sin\theta} . \mathbf{\hat {\bf \phi}}  + \nu D^2 v',
\label{NS}
\end{equation}
where $D^2$ is the operator
$\frac{\partial^2}{\partial r^2}+\frac{2}{r}\frac{\partial}{\partial r}+\frac{1}
{r^2\sin\theta}(\frac{\partial}{\partial\theta}(\sin\theta\frac{\partial}{\partial
\theta})-\frac{1}{\sin\theta})$  and $\mu_0$ is the induction constant.
The source of the sole nonlinearity in the dynamo equation 
is the feedback of the azimuthal component of the
Lorentz force (Eq.\ (\ref{NS})), which modifies only slightly the underlying imposed rotation law, 
but thus limits the magnetic fields at finite amplitude.
The assumption of axisymmetry allows the field ${\bf B}$ to be split simply
into toroidal and poloidal parts,
${\bf B}={\bf B}_T+{\bf B}_P = B\hat\phi +\nabla\times A\hat\phi$,
and Eq.\ (\ref{mfe}) then yields two scalar equations for $A$ and $B$.
Nondimensionalizing in terms of the solar radius $R$ and time $R^2/\eta_0$,
where $\eta_0$ is the maximum value of $\eta$, and
putting $\Omega=\Omega^*\tilde\Omega$, $\alpha=\alpha_0\tilde\alpha$,
$\eta=\eta_0\tilde\eta$, ${\bf B}=B_0\tilde{\bf B}$ and $v'= \Omega^* R\tilde v'$,
results in a system of equations for $A,B$ and $v'$. The
dynamo parameters are the 
two magnetic Reynolds numbers $R_\alpha=\alpha_0R/\eta_0$ and
$R_\omega=\Omega^*R^2/\eta_0$, and the turbulent Prandtl number
$P_r=\nu_0/\eta_0$. 
$\Omega^*$ is the solar surface equatorial angular velocity and
$\tilde\eta=\eta/\eta_0$. 
Thus $\nu_0$ and $\eta_0$ are the turbulent magnetic 
diffusivity and viscosity respectively,
$R_\omega$ is fixed when $\eta_0$ is determined (see Sect.~\ref{res}),
but the value of $R_\alpha$ is more uncertain.
The density $\rho$ is assumed to be uniform.

When attempting to model astrophysical systems, boundary conditions
are
often rather ill--determined. We try to make physically motivated choices.
For our inner boundary conditions we chose $B=0$, ensuring angular
momentum conservation, and an overshoot--type condition on ${\bf B}_P$
(cf.\ Moss \& Brooke 2000).
At the outer boundary, we used an open
boundary condition $\partial B/\partial r = 0$ on $B$ and
vacuum boundary conditions for ${\bf B}_P$. The motivation for this
is that the surface boundary condition is ill--defined, and there is some 
evidence that the more usual $B=0$ condition may be inadequate.
This issue has recently been discussed at length by Kitchatinov et al.\
(2000), who derive `non--vacuum' boundary conditions on both
$B$ and ${\bf B}_P$.

Equations (\ref{mfe}) and (\ref{NS}) were solved using the code 
described in Moss \& Brooke (2000) (see also
Covas et al.\ 2000) together with the above boundary conditions,
over the range $r_0\leq r\leq1$, $0\leq\theta\leq \pi$.
We set  $r_0=0.64$; with
the solar convection zone proper being thought to occupy the region $r \gta 0.7$,
the region $r_0 \leq r \lta 0.7$ can be thought of as an overshoot 
region/tachocline.
In the following simulations we used a mesh resolution of $61 \times 101$ 
points, uniformly distributed in radius and latitude respectively.

In this investigation, we took $\Omega_0$ to be given in
$0.64\leq r \leq 1$ by an interpolation on the MDI data 
obtained from 1996 to 1999 (Howe et al.\ 2000a).
For $\alpha$ we took $\tilde\alpha=\alpha_r(r)f(\theta)$,
where
$f(\theta)=\sin^2\theta\cos\theta$
(cf.\ R\"udiger \& Brandenburg 1995)
and 
$\alpha_r=1$ for $0.7 \leq r \leq 0.8$
with cubic interpolation to zero at $r=r_0$ and $r=1$,
with the convention that $\alpha_r>0$
and $R_\alpha < 0$. Also, in
order to take into account the 
likely decrease in the turbulent diffusion coefficient $\eta$ 
in the overshoot region, we allowed a simple
linear decrease from $\tilde\eta=1$ at $r=0.8$
to $\tilde\eta=0.5$ in $r<0.7$.

\section{Results}
\label{res}

We calibrated our model so that near marginal excitation 
the cycle period was about 22 years. 
This determined $R_\omega=44000$, corresponding to
$\eta_0\approx 3.4 \times 10^{11}$ cm$^2$ sec$^{-1}$,
given the known values of $\Omega^*$ and $R$.
The first solutions to be excited in the linear theory
are limit cycles with odd (dipolar) parity with respect to the equator, with 
marginal dynamo number
$R_\alpha \approx -2.23$. The even parity (quadrupolar) solutions
are also excited at a similar marginal 
dynamo number of  $R_\alpha \approx -2.24$.
It is plausible that the turbulent Prandtl number be of order unity,
and we set $P_r=1$.
For the parameter range that we investigated, the even parity solutions
are nonlinearly stable. Given that the Sun is observed to be close to
an odd (dipolar) parity state, and that previous experience shows that small
changes in the physical model can cause a change between odd and even parities
in the stable nonlinear solution, we chose to impose dipolar parity on our
solutions.

\begin{figure}[!htb]
\centerline{\def\epsfsize#1#2{0.387#1}\epsffile{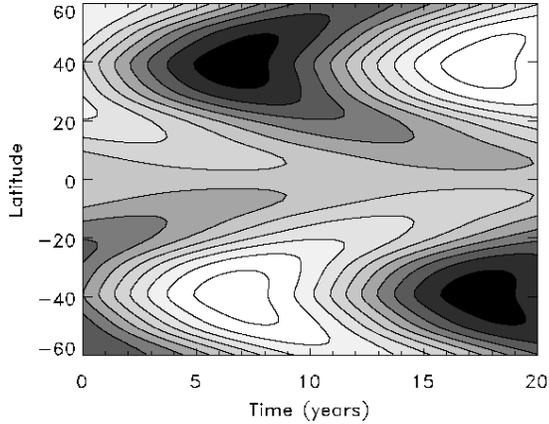}}
\caption{\label{Comega=44000.P=-1.0.butterfly_bp_60.degrees}
Butterfly diagram of the toroidal component of the
magnetic field $\vec{B}$ at fractional radius $r=0.95$.
Dark and light shades correspond to positive and negative values of
$B_\phi$ respectively.
Parameter values
are $R_{\alpha}=-3.0$, $P_r=1.0$ and ${R_{\omega}}=44000$.
}
\end{figure}

\begin{figure}[!htb]
\centerline{\def\epsfsize#1#2{0.387#1}\epsffile{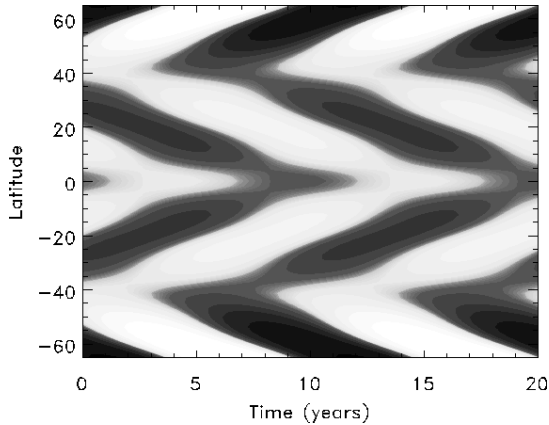}}
\caption{\label{Comega=44000.P=-1.0.velocity.R=0.99}
Variation of rotation rate with latitude and time from
which a temporal average has been subtracted to reveal the
migrating banded zonal flows, taken at fractional radius $r=0.99$. 
Darker and lighter regions
represent positive and negative deviations
from the time averaged background rotation rate.
Parameter values are as in Fig.\ {\protect
\ref{Comega=44000.P=-1.0.butterfly_bp_60.degrees}}.}
\end{figure}

With these parameter values, we found that this model,
with the underlying zero order angular velocity
chosen to be consistent with the recent (MDI) helioseismic
data, is capable of
producing butterfly diagrams which are in qualitative agreement
with the observations. An example is shown in
Fig.\ \ref{Comega=44000.P=-1.0.butterfly_bp_60.degrees}.
(The polar branch is a little too strong, but this feature can be 
weakened by adjusting the latitudinal dependence of $\alpha$ (see
also Covas et al.\ 2000).)
The model can also successfully produce torsional oscillations (see Fig.\
\ref{Comega=44000.P=-1.0.velocity.R=0.99})
that penetrate into the convection zone, similar to those deduced from
recent helioseismic data (Howe et al.\ 2000a) and studied in
Covas  et al.\ (2000). We note, however, that an additional interesting
feature of the present model is that the torsional oscillations
have larger and more realistic amplitudes near the surface, of the order 
of 1 nHz, much larger than was found previously using the boundary condition $B=0$
at the surface.

\begin{figure}[!htb]
\centerline{\def\epsfsize#1#2{0.387#1}\epsffile{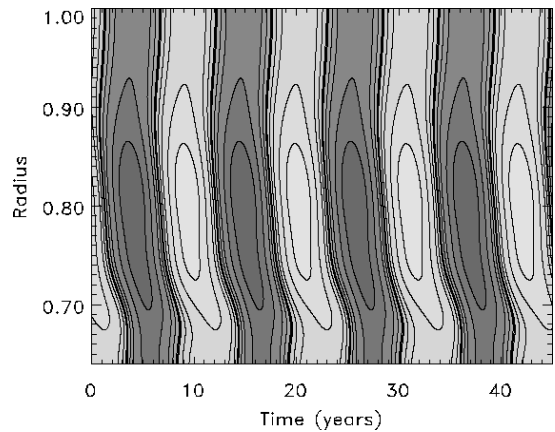}}
\caption{\label{perturbed.velocity.r.butterfly1} 
Radial contours of the angular velocity residuals $\delta \Omega$
as a function of time for a cut at latitude $30^{\circ}$.
Parameter values
are  $R_{\alpha}=-3.0$, $P_r=1.0$,
$R_{\omega}=44000$. Note how the torsional oscillations are very coherent
from top to the base of the dynamo region showing that only one period is
present.
}
\end{figure}

\begin{figure}[!htb]
\centerline{\def\epsfsize#1#2{0.387#1}\epsffile{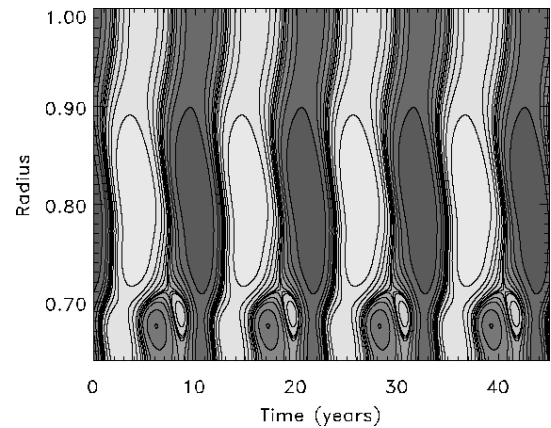}}
\caption{\label{perturbed.velocity.r.butterfly2} 
Radial contours of the angular velocity residuals $\delta \Omega$
as in Fig.\ {\protect \ref{perturbed.velocity.r.butterfly1}} for
$R_{\alpha}=-7.0$. Note the emergence of spatiotemporal fragmentation
towards the bottom of the convective zone, resulting in different periodicities there.
}
\end{figure}

\begin{figure}[!htb]
\centerline{\def\epsfsize#1#2{0.387#1}\epsffile{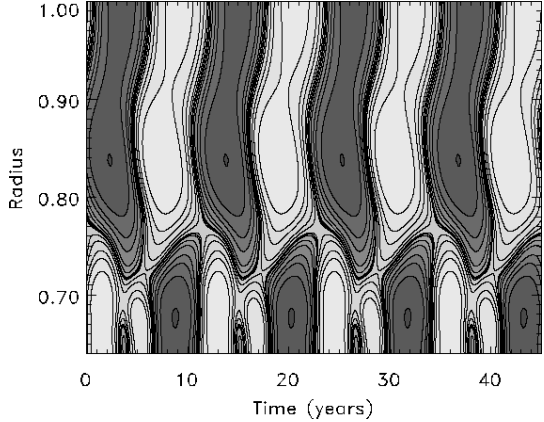}}
\caption{\label{perturbed.velocity.r.butterfly3} 
Radial contours of the angular velocity residuals $\delta \Omega$
as in Fig.\ {\protect \ref{perturbed.velocity.r.butterfly1}} for
$R_{\alpha}=-10.0$. 
}
\end{figure}

We found that the model is also capable of producing 
spatiotemporal fragmentation,
near the base of the convection zone,
hence resulting in oscillations
in the differential rotation 
with, for example, half the basic period.
To demonstrate this, we have plotted in Figs.\ \ref{perturbed.velocity.r.butterfly1}--\ref{perturbed.velocity.r.butterfly3}
the radial contours of the angular velocity residuals $\delta \Omega$
as a function of time for a cut at latitude $30^{\circ}$, for several
values of $R_{\alpha}$. As can be seen,
for smaller values of $R_{\alpha}$ (Fig.\ \ref{perturbed.velocity.r.butterfly1}),
we find torsional oscillations with the same period at the top
and the bottom of the convection zone.
As $R_{\alpha}$ is increased (Figs.\ \ref{perturbed.velocity.r.butterfly2} 
and \ref{perturbed.velocity.r.butterfly3}), a spatiotemporal
fragmentation occurs
near the base of the convection 
zone,
resulting in oscillations in the 
differential rotation with half the period
of the oscillations near the top. For still higher values of $R_{\alpha}$,
the temporal variations in the differential rotation at
the base of the convection zone start to become non-periodic, which might be 
of relevance if the failure of Antia 
\& Basu (2000) to find shorter period oscillations
near the bottom of the convection zone
should turn out to be correct.
We have also checked that the 
butterfly diagrams do not fragment
and keep the same period independently of the depth and $R_\alpha$ value,
continuing to resemble Fig.\ \ref{Comega=44000.P=-1.0.butterfly_bp_60.degrees}.

\begin{figure}[!htb]
\centerline{\def\epsfsize#1#2{0.440#1}\epsffile{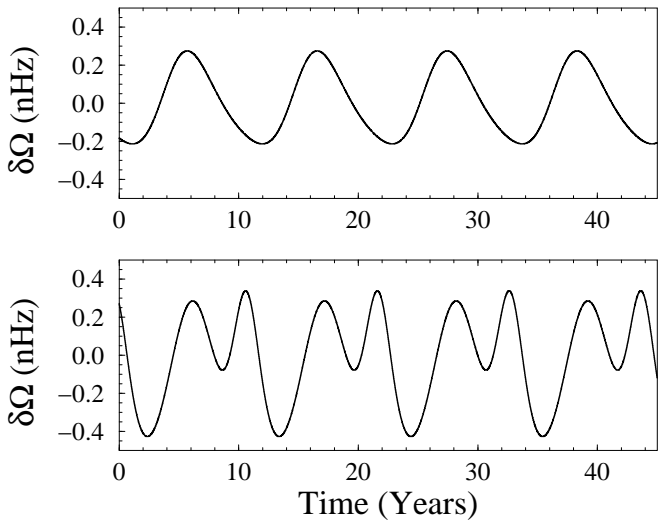}}
\caption{\label{double.period} 
`Period halving' at $r=0.68$ and latitude $30^{\circ}$.
The panels correspond, from left to right, to $R_{\alpha}$ values $-3.0$ and
$-7.0$ respectively, and display
increasing relative amplitudes of the secondary oscillations.
Remaining parameters values
are as in Fig.\ {\protect \ref{perturbed.velocity.r.butterfly1}}.
}
\end{figure}

This fragmentation is made more transparent
in Fig.\ \ref{double.period} which shows 
the temporal
oscillations in the angular velocity residuals $\delta \Omega$
at a fixed point,
as ${R_{\alpha}}$ is increased, illustrating the presence of 
period halving.

\section{Discussion}
We have proposed spatiotemporal fragmentation as a natural
mechanism for producing different types of variation in the differential rotation
at the top and the bottom of the convection
zone.
To demonstrate the occurrence of such behaviour, 
we have studied a 
solar dynamo model, with a semi--open outer boundary
condition, calibrated to have the correct
cycle period, with a mean rotation law given by recent 
helioseismic observations. 
We note in passing that in a few simulations performed   
with the boundary condition $B=0$ at the surface,
we have not so far found this phenomenon, although we
cannot yet make a definitive statement on this point.
In addition to producing
butterfly diagrams in qualitative
agreement with those that are observed,
as well as torsional oscillations that
penetrate into the convection zone, we have shown that this model
can also produce spatiotemporal fragmentation, resulting
in different oscillatory modes of behaviour near the 
top and the bottom of the convection zone.

We emphasize that the main aim of this letter is
to propose a mechanism that can be
expected to operate in general nonlinear dynamo settings,
and which is capable of producing multiple periods and/or non-periodic
oscillations in parts of the convective zone.
The specific results given here, such as the single period halving,
are based on a particular dynamo model which inevitably
includes many simplifying assumptions, not least of which is that the density is
uniform.
(It is unclear how the inclusion of a radial dependence $\rho(r)$ would 
affect our results --- we note that current solar dynamo models 
commonly take a uniform density.)
We expect that the mechanism is of quite general applicability, and so
it is plausible that a more sophisticated model might exhibit further
bifurcations, 
thus producing different
reduced periods and oscillatory regimes.  
It may also be useful to bear in mind in this connection
that three period halvings would result in $11$ years$/2^3 \sim 1.3$ years!
We shall return to a more detailed study of the underlying
dynamics as well as a quantitative study
of different dynamo models elsewhere.
We have chosen $P_r=1$ in order to obtain larger
amplitude torsional oscillations near the surface. We have checked that 
fragmentation still occurs at smaller values of 
$P_r$.

Inevitably the uncertainties associated with the inversion of the helioseismic data
so deep in the convection zone are quite large. Thus we believe that
the mechanism discussed here may, by demonstrating what modes of 
dynamical behaviour are theoretically possible,
act as a conceptual aid in interpreting current 
and further observations.
\begin{acknowledgements}
We would like to thank R. Howe and M. Thompson for
many useful discussions regarding the helioseismic observations
and for providing us with the time average of the solar
rotation. 
EC is supported by a PPARC fellowship.
RT would like to thank R. Brandenberger and
R. Unruh for their hospitality
during a visit to the University 
of British Columbia, where some of this
work was done.
\end{acknowledgements}

\end{document}